\documentclass[prb,twocolumn,showpacs,floatfix,superscriptaddress]{revtex4}
\usepackage{epsfig,amssymb,amsmath,subfigure}

\newcommand*{\atow}[0]{$\alpha$$\to$$\omega$}
\newcommand*{\eps}[0]{\varepsilon}
\newcommand*{\epsmax}[0]{\eps_{\text{max}}}
\newcommand*{\deltamax}[0]{\delta_{\text{max}}}
\newcommand*{\dcom}[0]{\delta_{\text{COM}}}
\newcommand*{\drms}[0]{\delta_{\text{RMS}}}
\newcommand*{\dca}[0]{d_{\text{ca}}}
\newcommand*{\Ecut}[0]{E_{\text{cut}}}
\newcommand*{\ohmms}[0]{\textsc{ohmms}}

\newcommand*{\vasp}[0]{\textsc{vasp}}
\newcommand*{\Vasp}[0]{\textsc{Vasp}}
\newcommand*{\comsubs}[0]{\textsc{comsubs}}
\newcommand*{\abinit}[0]{\textit{ab~initio}}
\newcommand*{\Abinit}[0]{\textit{Ab~initio}}
\newcommand*{\NEBVASP}[0]{\abinit\ NEB}

\newcommand*{\et}[0]{\textit{et~al.}}

\newcommand*{\x}{\times}

\newcommand*{\tr}[1]{#1^\mathrm{T}}
\newcommand*{\inv}[1]{#1^{-1}}

\newcommand*{\crys}[1]{\textbf{#1}}
\newcommand*{\term}[1]{{\em #1}}
\newcommand*{\pgs}[0]{pathway generation and sorting}

\newcommand*{\be}[0]{\begin{equation}}
\newcommand*{\ee}[0]{\end{equation}}
\newcommand*{\beu}[0]{\begin{equation*}}
\newcommand*{\eeu}[0]{\end{equation*}}
\newcommand*{\ba}[0]{\begin{array}}
\newcommand*{\ea}[0]{\end{array}}

\newcommand*{\entry}[1]{\multicolumn{1}{c}{\underbar{#1}}}

\begin{document}

\title{Systematic pathway generation and sorting in martensitic 
  transformations:\\ Titanium $\alpha$ to $\omega$}

\author{D. R. Trinkle}
\affiliation{Materials and Manufacturing Directorate, Air Force Research Laboratory, Wright Patterson Air Force Base, Dayton, Ohio 45433-7817}
\author{D. M. Hatch}
\affiliation{Brigham Young University, Provo, UT 84602}
\author{H. T. Stokes}
\affiliation{Brigham Young University, Provo, UT 84602}
\author{R. G. Hennig}
\affiliation{The Ohio State University, Columbus, OH 43210}
\author{R. C. Albers}
\affiliation{Los Alamos National Laboratory, Los Alamos, NM 87545}

\date{\today}

\begin{abstract}
Structural phase transitions are governed by the underlying atomic
transformation mechanism; martensitic transformations can be separated
into strain and shuffle components.  A systematic pathway generation and sorting
algorithm is presented and applied to the problem of the titanium $\alpha$ to
$\omega$ transformation under pressure.  In this algorithm, all
pathways are constructed within a few geometric limits, and
efficiently sorted by their energy barriers.  The geometry and
symmetry details of the seven lowest energy barrier pathways are
given.  The lack of a single simple geometric criterion for
determining the lowest energy pathway shows the necessity of atomistic
studies for pathway determination.
\end{abstract}

\pacs{05.70.Fh, 61.50.Ks, 64.70.Kb, 81.30.Kf}

\maketitle

Martensitic transformations are first order, diffusionless,
displacive, athermal crystal structure transformations.  They deform
the lattice and change the shape of the crystal while the relative
motion of the atoms is small compared to the nearest neighbor
distances.\cite{Haasen96} The lattice deformation of martensitic
transformation results in a characteristic orientation relationships:
Specific vectors and planes in the initial lattice are transformed to
specific vectors and planes in the final lattice.  The orientation
relations can be measured experimentally for many materials and
provides constraints on the possible atomistic pathways of the
transformation.  Furthermore, martensitic transformations proceed near
the speed of sound, complicating a direct observation of the atomic
motion.\cite{Martensite} Despite these conditions on the atomic
motion, the atomistic pathway for many martensitic
transformations---the precise motion of each atom during the
transformation---remains unknown.

The problem of finding the most likely pathway for a transformation
reduces to generating a relevant subset of possible pathways and
sorting by their energy barriers.  There are infinitely many unique
ways to transform one crystal continuously into another.  Only a
finite subset of these, however, results in small strain and small
atomic motion.  Our ``\pgs'' algorithm produces the set of pathways
with small total shape change---strain---and small atomic
motion---shuffle.  The pathway generation is related to
\comsubs.\cite{Stokes02,COMSUBS}  From this set, we use the energy
barrier of each transformation pathway to determine the most likely
pathway.

Titanium's room temperature $\alpha$ phase (hcp) transforms to the
high pressure $\omega$ phase (3-atom hexagonal cell) at approximately
2--9~GPa.\cite{Sikka82}  This transformation lowers toughness and
ductility in Ti alloys.  Like all martensitic transformations, 
\atow\ occurs at the atomic level, so
experiments\cite{Silcock58,Usikov73,Kutsar73,Vohra81,Gray92,Dammak99}
can only provide some limits on possible mechanisms; moreover, without
a systematic algorithm for generation, it is impossible to determine
the lowest energy barrier pathway.  These problems meant that the
mechanism for the \atow\ transformation was only recently elucidated
by Trinkle~\et\cite{Trinkle03}

To apply the \pgs~algorithm to Ti~\atow, we proceed in several steps.
We begin by describing the notation used.  We then solve half of the
pathway generation problem by determining the strains for each
pathway.  Next, we solve the other half of the pathway generation
problem by determining internal relaxations needed for each pathway.
Finally, we combine the pathway generation methods with the energy
barrier evaluation methods into a general algorithm for producing all
relevant pathways and sorting by energy barrier, and apply it to the
problem of the Ti \atow\ transformation.\cite{Trinkle03}  Finally, the
geometry, symmetry, and energy barriers of the seven lowest energy
barrier pathways are given.

\section{Introduction}

\subsection{Definition of pathway}

A \term{pathway} between two infinite periodic crystals \crys{a} and
\crys{b} is a reversible mapping between the atoms in crystal \crys{a}
and \crys{b}.  Once each atom in \crys{a} is identified with an atom
in \crys{b} we continuously transform from one crystal into another;
moreover, the position of each atom during this transformation is
calculated to determine the energy barrier of the transformation.  We
separate the problem of finding mappings between the
endpoints---``pathway generation''---from following the transformation
from \crys{a} to \crys{b} to determine the {\em energy barrier} for finding
the lowest energy pathway.  Once energy barriers are known, we choose
the lowest energy barrier pathway---the ``sorting'' of the pathways.

Pathways for martensitic phase transformations can be separated into
\term{strain}---a global shape change---and
\term{shuffle}---small atomic relaxations.  The strain results in a
macroscopic change of lattice vectors $\vec a_i$.  The shuffles produce
changes in the atomic basis $x_j$.

We restrict ourselves to the mappings that are periodic for the entire
transformations.  This restricts us to homogeneous pathways from
infinite crystal \crys{a} to \crys{b}.  By enforcing periodicity, we
consider only certain transformation strains.  The shuffles for each
pathway are periodic; this periodicity allows for a finite search of
possible pathways.  Enforcing homogeneity makes the homogeneous energy
barrier the relevant ``figure of merit'' for each pathway.

We scale the volumes of our crystals so that the volume per atom in
\crys{a} is equal to that in \crys{b}.  This simplifies the
calculation of the strain.  For an appreciable volume change, a
volumetric strain can be added to the resulting pathways.

\subsection{Notation}

A crystal \crys{a} has lattice vectors written as a matrix $[a]$ of
column vectors. The volume of the unit cell is $\det [a]$; moreover,
for the lattice vectors $\{\vec a_i\}$ to be linearly independent,
$\det [a] \ne 0$.  The atomic basis vectors $x_j$ are column vectors
in unit cell coordinates; we can represent all cartesian points in the
infinite lattice using integer column vectors $k$, $\vec R(k; j) =
[a](x_j + k)$.  In general, $[a]$ and $[b]$ will represent the lattice
vectors of crystals \crys{a} and \crys{b}, respectively; we will also
use $[a]$ to denote the lattice.

\label{ch:method:sublattice}

A \term{sublattice} $[A]$ of a lattice $[a]$ is a lattice where every
point in $[A]$ is also in $[a]$.  The lattice vectors $[A]$ are called
the \term{supercell}.  In order for $[A]$ to be a sublattice of $[a]$,
there must exist an non-singular integer matrix $[n]$ such that $[A] =
[a][n]$.  This condition guarantees that any point in $[A]$ is also a
point in $[a]$.  The \term{size} of the supercell $[A]$ is the ratio
of the number of points in $[a]$ to $[A]$; it is $\det [n]$.
Generally, we will write $[A]$ and $[B]$ for a supercell of $[a]$ and
$[b]$, respectively; the integer matrix $[n]$ will relate $[a]$ to $[A]$,
and $[m]$ will relate $[b]$ to $[B]$.

Atoms in crystal \crys{a} are mapped into new positions in the crystal
\crys{A}.  The basis for \crys{A} is $\{u_i^a\}$; each atom in \crys{A}
is identified with some atom in \crys{a}: $[A]u_i^a = [a](x_j + k)$,
where $k$ is a column vector of integers.  This means that $u_i^a =
\inv{[n]}(x_j + k)$ for some integer vector $k$.  We require that each
component of $u_i^a$ be in $[0,1)$, so there are exactly $\det [n]$
possible vectors $k$.  This means that if \crys{a} has $N_a$ atoms,
then \crys{A} will have $N_a\det [n]$ atoms.

A \term{pathway} is described as a supercell pair $[A]$ and $[B]$ with
$N$ atom positions $u_i$ and shuffles $\delta_i$.  Each supercell
relates to its crystal by $[A](u_i) = [a](x_j^a + k)$ and
$[B](u_i+\delta_i) = [b](x_j^b + k') + (\text{constant shift})$, where
$k$ and $k'$ are integer vectors and the constant shift is a uniform
translation of every point in the lattice.  These equations connect
the atom positions $u_i$ and shuffles $\delta_i$ to the underlying
crystals \crys{a} and \crys{b}.  Moreover, the supercells $[A]$ and
$[B]$ are related by a symmetric strain tensor $\eps$ and a rotation
matrix $\Phi$,
\be
\label{eqn:method:strain}
\eps [A] = \Phi [B].
\ee
The strain $\eps$ changes the lattice vectors $[A]$ into those of
$[B]$ after a rotation.  This corresponds to a change in the lattice
vectors $[a]$ since $[A] = [a][n]$.  Moreover, the strain is
determined {\em entirely} by the supercells.

These equations allow us to write each \term{pathway} as $P([n],
[m]; u_i, \delta_i)$, where $[A] = [a][n]$ and $[B] = [b][m]$.  The
advantage of working in terms of the integer matrices $[n]$ and $[m]$
is the convenience in enumerating all possible supercells of interest.
We will often talk about a \term{supercell pair} $[n]$ and $[m]$ and
write $P([n], [m]; *)$ to indicate that atom positions and shuffles
have not yet been determined for this pathway.

We determine the strain, orientation relations, distance that atoms
move, distance of closest approach to other atoms, and common subgroup
from $P([n], [m]; u_i, \delta_i)$.  The strain $\eps$ is determined
purely by the supercell pair $P([n], [m]; *)$.  The \term{orientation
relations}---which vectors and planes in the lattice $[a]$ are
transformed into vectors and planes in the lattice $[b]$---are also
determined from $P([n], [m]; *)$.  The shuffle information is
sufficient to determine a continuous linear simultaneous
transformation from \crys{a} to \crys{b}; we use this transformation
to calculate the distance that each atom moves and the distance of
closest approach.

We determine whether two pathway representations $P$ and $P'$ are
different by comparing their supercell pairs and their shuffles. At
each step, equivalent pathway representations are removed.  The
details of these comparisons are in
Section~\ref{ch:method:unique-pairs} for the supercell pairs and
Section~\ref{ch:shuffles:unique} for the shuffles.

\section{Solving the strain problem}

The first step in enumerating possible pathways is to find the set of
supercell pairs $P([n], [m]; *)$.  We find all supercell pairs of a
given size $N$ where the strain $\eps$ is limited by a cutoff
$\epsmax$.  We begin by enumerating possible unique sublattices of a
given size, and reduce this list by the lattice symmetry.  We then
revisit Eq.~(\ref{eqn:method:strain}) and solve for the strain $\eps$
and rotation $\Phi$.  We restrict strains to a cutoff $\epsmax$ to
produce a finite search algorithm enumerating all possible supercell
pairs.  Some aspects of this problem were originally considered in
Lomer's calculation of the orientation relations in U
$\beta$$\to$$\alpha$.\cite{Lomer56}

\subsection{Unique sublattices}
\label{ch:method:unique-sub}

We say two lattices \crys{x} and \crys{y} with lattice vectors $[x]$
and $[y]$, respectively, are \term{equivalent} if every point in $[x]$
is a point in $[y]$ and vice versa.  This mirrors the definition of
sublattice from Section~\ref{ch:method:sublattice}; i.e., $[x]$ and
$[y]$ are equivalent if and only if $[x]$ is a sublattice of $[y]$ and
$[y]$ is a sublattice of $[x]$.  This means there are two non-singular
integer matrices $[i]$ and $[j]$ such that $[x] = [y][i]$ and $[y] =
[x][j]$.  Combining these equations shows that $[i]$ and $[j]$ must be
\term{unimodular}; that is, $\det [i] = \pm 1$ and $\det [j] = \pm
1$.  This gives another condition for equivalence: There exists an
integer unimodular matrix $[l]$ such that $[x] = [y][l]$.

We can apply this definition to supercells as well; two supercells,
given by $[n]$ and $[n']$ are equivalent if and only if there exists
an integer unimodular matrix $[l]$ such that $[n] = [n'][l]$.  This
also means that equivalent supercells must be of the same size; i.e.,
$|\det [n]| = |\det [n']|$.  We use unimodular matrices to iterate
over possible supercell representations for a sublattice; this is
equivalent to forming alternate supercells that have the same size.

From each set of equivalent supercells, we pick one representative
called $[\bar n]$; all other members of the set can be generated using
integer unimodular matrices $[l]$ as $[\bar n][l]$.  The
representative sublattices are given by an upper-triangular matrix
$[\bar n]$, where $\bar n_{ij}\ge0$ for all $i,j$, $\bar n_{ij} < \bar
n_{ii}$ for $j>i$, and $\prod_{i=1}^d \bar n_{ii} = N$, where $N$ is
the size of the sublattice $\det [\bar n]$.  It is straightforward to
show that for a given $N$ there are a finite set of supercells in this
form.  If two matrices $[\bar n]$ and $[\bar n']$ have the above form,
they are equivalent if and only if they are equal (see Appendix~D of
[\onlinecite{TrinkleThesis}]).

Using the form for representative supercells we enumerate all possible
unique sublattices for $[a]$ and $[b]$.  The set of unique sublattices
for $[a]$ is $\{[\bar n]\}$ and for $[b]$ is $\{[\bar m]\}$.  We
restrict ourselves to pathways of a given size $N$; this requires
that $N_a \det [\bar n] = N = N_b \det [\bar m]$, where $N_a$ and
$N_b$ are the number of atoms in crystal \crys{a} and \crys{b},
respectively.

To further reduce the set of representative supercell, consider the
set of all symmetry operations $G_a$ on $[a]$ where at least one
lattice point is mapped to itself; this is the \term{point group} of
\crys{a}.  We can represent each member of the point group $G_a$ with a
matrix $g_a$ that operates on the cartesian coordinates of $[a]$.  In
order to be a symmetry element, $g_a$ must map the lattice $[a]$ back
onto itself: There exists an integer unimodular matrix $\bar g_a$ such
that $g_a [a] = [a] \bar g_a$.  Since the lattice \crys{a} is left
invariant under the operation of $g_a$, two initially {\em different}
sublattices \crys{A} and \crys{A'} may be {\em equivalent} by $g_a$.

To determine the set of equivalent sublattices of size $N$ for a given
lattice \crys{a}, we start with the set of unique representations
$\{[\bar n]\}$ and reduce it by symmetry.  We say that two of our
(initially unique) representations $[\bar n]$ and $[\bar n']$ are
equivalent if there exists a symmetry element $g_a$ (with integer
unimodular matrix $\bar g_a$) and some integer unimodular matrix $[i]$
such that $\bar g_a [\bar n] = [n'] = [\bar n'][i]$.  Thus, our
initial list of unique sublattices of size $N$ may be further reduced
by the symmetry of lattice \crys{a}.  Such a reduction can also be
applied to the sublattices of lattice \crys{b}.

\subsection{Calculating and limiting the strain}
\label{ch:method:strain}

Given two sets of sublattice representatives $\{[\bar n]\}$ and
$\{[\bar m]\}$ for \crys{a} and \crys{b}, respectively, we find all
strains that transform from one sublattice into the other.  We limit
the allowed strain by a maximum cutoff $\epsmax$, which produces a
finite list of possible supercell pairs.  We begin by solving the
strain Eq.~(\ref{eqn:method:strain}) for the general case.  We
then limit the allowed strain, and translate that limit into a subset
of allowed supercell pairs.

We take a supercell representative $[\bar n]$ for \crys{a} and $[\bar
m]$ for \crys{b} and substitute into Eq.~(\ref{eqn:method:strain})
to get an equation for the possible strains from $[\bar n]$ to $[\bar
m]$.  The supercells are $[A] = [a][\bar n][i]$ for some integer
unimodular matrix $[i]$, and $[B] = [b][\bar m][j]$ for some integer
unimodular matrix $[j]$.  The equation for $\eps$ and $\Phi$ is $\eps
[a][\bar n][i] = \Phi [b][\bar m][j]$, which can be simplified by
right multiplying by $\inv{[i]}$, 
\be
\label{eqn:method:eps}
\eps [a][\bar n] = \Phi [b][\bar m][l],
\ee
where $[l] = [j]\inv{[i]}$ is an integer unimodular matrix.

To find all possible strains, we will solve
Eq.~(\ref{eqn:method:eps}) for all integer unimodular matrices
$[l]$ to find the supercell pairs $P([\bar n], [\bar m][l]; *)$.  We
right multiply Eq.~(\ref{eqn:method:eps}) by $\inv{[\bar
n]}\inv{[a]}$ to get $\eps = \Phi [b][\bar m][l]\inv{[\bar
n]}\inv{[a]}$.  We define the matrix $C([l]) = [b][\bar
m][l]\inv{[\bar n]}\inv{[a]}$; then $\eps = \Phi C([l])$.  The strain
tensor $\eps$ is symmetric, though $C([l])$ and $\Phi$ are not.  We
left multiply each side of the equation by its transpose to get
\be
\label{eqn:method:square}
\begin{split}
\tr{\eps}\eps &= \tr{C([l])}\tr{\Phi}\Phi C([l]), \\
\eps^2 &= \tr{C([l])}C([l]),
\end{split}
\ee
where the second equality is true because $\eps$ is symmetric and
$\Phi$ is a rotation matrix, so $\tr{\Phi}=\inv{\Phi}$.  Thus, $\eps$
is the square root of $\tr{C([l])}C([l])$, a symmetric matrix.  We
calculate the square root by diagonalizing $\tr{C([l])}C([l])$ and
writing it as $\tr{\Theta} c \Theta$ where $\Theta$ is a rotation
matrix and $c$ is a diagonal matrix of the eigenvalues.  Then $\eps =
\tr{\Theta}\sqrt{c}\Theta$, and $\sqrt{c}$ are the diagonal strain
elements of $\eps$; because there is no volumetric change, the product
of the diagonal strain elements are unity.  Once $\eps$ is known,
$\Phi = \eps [a][\bar n]\inv{[l]}\inv{[\bar m]}\inv{[b]}$.  This
method may also be considered an extension of the method of magic
strains.\cite{VandeWaal90}

Though every integer unimodular matrix $[l]$ has an associated strain
tensor $\eps$, most of these will be large.  We define a strain
limitation in terms of the cutoff $\epsmax>1$; a strain $\eps$ is
within our cutoff if for all non-zero vectors $\vec v$,
\be
\label{eqn:method:epsmax}
\inv{\epsmax}|\vec v| \le |\eps \vec v| \le \epsmax |\vec v|.
\ee
This condition is equivalent to requiring that all the eigenvalues of
our strain matrix are between $\inv{\epsmax}$ and $\epsmax$.  We
consider only those $[l]$ that will produce a strain within our
$\epsmax$ cutoff.

Eq.~(\ref{eqn:method:strain}) relates each supercell lattice
vector $\vec A_i$ to a supercell lattice vector $\vec B_i$; this
translates Eq.~(\ref{eqn:method:epsmax}) to a condition on each
$\vec B_i$.  We substitute $\vec A_i$ for $\vec v$ in
Eq.~(\ref{eqn:method:epsmax}) noting that $\eps \vec A_i$ is
$\vec B_i$, so that for each $i$,
\be
\label{eqn:method:epsmax-latt}
\inv{\epsmax}|\vec A_i| \le |\vec B_i| \le \epsmax |\vec A_i|.
\ee
Thus, if $\{\vec B_i\}$ doesn't satisfy
Eq.~(\ref{eqn:method:epsmax-latt}), then $\eps$ will {\em never}
satisfy Eq.~(\ref{eqn:method:epsmax}).  Hence, only $\vec B_i$
which lie in the annulus between $\inv{\epsmax}|\vec A_i|$ and
$\epsmax|\vec A_i|$ need to be considered; this is a finite list for
each $\vec B_i$.

We translate Eq.~(\ref{eqn:method:epsmax-latt}) into a condition on
the allowed unimodular integer matrices $[l]$.  For each $i$, we
construct the (finite) set of integer vectors $\{\lambda^{(i)}_j\}$
such that $B_i = [b][\bar m]\lambda^{(i)}_j$ satisfies
Eq.~(\ref{eqn:method:epsmax-latt}).  This set will be different for
each $i$, as well as for each supercell $[\bar n]$ and $[\bar m]$.  To
construct all of the possible unimodular matrices $[l]$, we check to
see if the matrix $[l(j_1,j_2,j_3)] = \left( 
\lambda^{(1)}_{j_1}\;\lambda^{(2)}_{j_2}\;\;\lambda^{(3)}_{j_3} \right)$
has determinant 1.  If it does, we determine $\eps$ for the supercells
$[\bar n]$ and $[\bar m][l(j_1,j_2,j_3)]$, and make sure that
$\eps$ is within the cutoff $\epsmax$.  To efficiently construct the
$\{\lambda^{(i)}_j\}$ we choose representations $[\bar n]$ and $[\bar
m]$ that have the {\em shortest} possible lengths in cartesian
coordinates for unit cells $[a]$ and $[b]$, respectively.  This makes
both $|\vec A_i|$ as short as possible for each $[\bar n]$, and makes
the set of $\lambda^{(i)}_j$ to check as small as possible.

\subsection{Uniqueness of supercell pairings}
\label{ch:method:unique-pairs}

Two different supercell pairs $P([\bar n], [\bar m][l]; *)$
and $P([\bar n'], [\bar m'][l']; *)$ are equivalent if they produce
the same mapping from \crys{a} to \crys{b}.  This first requires that
$[\bar n]=[\bar n']$ and $[\bar m]=[\bar m']$ since supercell
representatives are equivalent if and only if they are equal.
However, $[l]$ and $[l']$ need not be equal to give equivalent
mappings if they produce equivalent strains from \crys{a} to \crys{b}
and vice versa.

Two strains $\eps$ and $\eps'$ on a crystal \crys{a} are
\term{equivalent} if for arbitrary elastic constants $C^a_{ij}$ the
elastic energies $U_a(\eps)$ and $U_a(\eps')$,
\beu
U_a(\eps) = \frac{1}{2}\sum_{i,j=1}^6 C^{a}_{ij} e_i e_j,
\eeu
are equal.  The strains $e_i$ are related to the strain matrix $\eps$ by
\beu
\eps = \left(\ba{ccc}
1+e_1 & \frac{1}{2}e_6 & \frac{1}{2}e_5\\
\frac{1}{2}e_6 & 1+e_2 & \frac{1}{2}e_4\\
\frac{1}{2}e_5 & \frac{1}{2}e_4 & 1+e_3
\ea\right).
\eeu
The symmetry of the elastic constants $C^a_{ij}$ is given
by the class\cite{Nye57} of \crys{a}; there are 11 unique
three-dimensional crystal classes, and 5 unique two-dimensional
crystal classes.  We perform the same test for the strains on crystal
\crys{b}.  This test takes into account the point group symmetry of
the lattices \crys{a} and \crys{b}.

As an example, consider the cubic crystal class; it has only three
unique elastic constants $C_{11}$, $C_{12}$, and $C_{44}$.  We write
the energy for a general strain $e_i$ as:
\begin{align*}
U_a(\eps) &= \frac{1}{2}C_{11}(e_1^2+e_2^2+e_3^2)
             + C_{12}(e_1e_2 + e_2e_3 + e_3e_1) \\
          &\quad + \frac{1}{2}C_{44}(e_4^2+e_5^2+e_6^2).
\end{align*}
To determine if two strains are truly identical, we check that the
three combinations $e_1^2+e_2^2+e_3^2$, $e_1e_2 + e_2e_3 + e_3e_1$,
and $e_4^2+e_5^2+e_6^2$ are the same for $\eps$ and $\eps'$.  Similar
expressions can be derived for the other crystal classes.

\section{Determining shuffles}

Each supercell pair $P([n], [m]; *)$ represents the strain of the
transformation; to complete the pathway description requires knowledge
of the local atomic motion: the atom positions $u_i$ and shuffles
$\delta_i$.  We find the shuffles using a systematic approach to
enumerate all possible mappings between the atoms in the supercell
$[A]$ and the atoms in $[B]$.  We use geometric restrictions to reduce
the set of possible pathways to those that require small atomic
motions and do not have artificially small approach distances during
the transformation.

\subsection{Populating cells}
\label{ch:method:populate}

For a given supercell pair $P([n], [m]; *)$, we populate the $[A]$
supercell with the atoms from \crys{a} and the $[B]$ supercell with
the atoms from \crys{b} using the formula from
Section~\ref{ch:method:sublattice}.  The atom positions $u^a_i$ in
$[A]$ are $u_i^a = \inv{[n]}(x^a_j + k)$ for some integer vector $k$;
similarly, $u_i^b = \inv{[m]}(x^b_j + k')$.  We restrict all of the
components of $u_i^a$ and $u_i^b$ to be in $[0,1)$, so there are
exactly $N$ atoms from each crystal.

The shuffle vectors $\delta_i$ connect $u_i^a$ to the atoms $u_j^b$ in
the periodic supercell as the strain $\eps$ connects $[A]$ to $[B]$.
Because of the periodicity of the supercells, the mapping can connect
an atom $i$ to an periodic image of an atom $j$ that is not inside the
original supercell.  Suppose we have our mapping, and each $u_i^a$ is
transformed into some $u_{j(i)}^b + h_j$ where $h_j$ is an integer
vector; our shuffle vectors $\delta_i$ are $\delta_i = u_{j(i)}^b +
h_j - u_i^a$.

Our problem then is to find the mapping for each atom $i$ in $[A]$ to
an atom $j$ in $[B]$.  Moreover, we look for mappings that result in
small atomic motions and do not bring two atoms close together, both
of which produce large energy barriers.

\subsection{Center of mass}
\label{ch:method:com}

We require that the shuffles do not change the center of mass of the
crystal during the transformation: $\sum_i \delta_i = 0$.  This is
enforced by shifting all of the atoms $u_j^b$ by a constant vector
$\dcom$.  Then, $\delta_i = u_{j(i)}^b + \dcom + h_j - u_i^a$.  We
solve for $\dcom$ by summing $\delta_i$ over all $N$ atoms
\beu
\begin{split}
\sum_i \delta_i = 0 & = \sum_i (u_{j(i)}^b + \dcom + h_j - u_i^a), \\
  & = \sum_j u_j^b - \sum_i u_i^a + N\dcom + \sum_j h_j.\\
\end{split}
\eeu
Then
\beu
\dcom = \frac{1}{N}\sum_i (u_i^a - u_i^b) - \frac{1}{N}\sum_j h_j.
\eeu
The only term in this equation that depends on the specific mapping is
the sum over integer vectors $h_j$.  Thus, every center of mass shift
$\dcom$ has the form
\be
\label{eqn:method:COM}
\dcom = \frac{1}{N}\sum_i (u_i^a - u_i^b) + \frac{1}{N}(k_1,k_2,k_3),
\ee
where $k_i=0\ldots N-1$.  We need not consider shifts larger than
this, as shifts by integer vectors only translate all of the atoms
$u_i^b$ from one periodic supercell image to another.

Because every mapping will have a center of mass shift given by
Eq.~(\ref{eqn:method:COM}), we systematically attempt each shift
$\dcom$ on the atoms $u^b_j$, and make the mappings from $u_i^a$ to
$u_j^b$ keeping only those where $\sum_i \delta_i = 0$.  We still
consider mappings where $u_i^a$ moves to $u_j^b + h_j$, but restrict
the $h_j$ components to all be either --1, 0, or 1.  This prevents an
atom from moving across an entire supercell.

The $N^3$ center of mass shifts $\dcom$ are reduced by considering
only shifts that stay within the unit cell of \crys{a} and
\crys{b}.  There are some shifts $k/N$ where $[m]k/N$ is all integer;
this corresponds to shifting all of the atoms in \crys{b} by a lattice
vector of $[b]$.  This will create no unique pathways, but rather
previously considered pathways with a permutation of the atom indices
$j$.  Similarly, some shifts correspond to shifting all of the atom by
a lattice vector of $[a]$; these need not be considered either.  We
reduce the set of possible $\dcom$ to those that are within a single
unit cell of $[a]$ (defined by $[n]$) and $[b]$ (defined by $[m]$).
Depending on the orientation of $[n]$ and $[m]$, this reduces the
number of center of mass shifts to somewhere between $N^3/(N_aN_b)$
and $N^3/\max(N_a,N_b)$.

\subsection{Mapping atoms for a given center of mass shift}
\label{ch:method:mapping}

For each supercell pair $P([n], [m]; *)$, we loop over the set of
possible center of mass shifts $\dcom$ and enumerate all possible
mappings for each center of mass shift.  This becomes a combinatoric
problem; to reduce the exponential scaling, we examine only mappings
where no atom moves ``far,'' given by a shuffle cutoff $\deltamax$.
We further reduce the set of possible mappings to those where atoms do
not approach closely and where the total shuffle for all atoms is
small.

We define a metric $d(x-y)$ between two vectors $x$ and $y$ in
supercell coordinates
\be
d(x-y) = \sqrt{\frac{1}{2} |[A](x-y)|^2 + \frac{1}{2} |[B](x-y)|^2}.
\label{eqn:mapping:distance}
\ee
This distance function is symmetric in the supercells $[A]$ and $[B]$,
is zero if and only if $x=y$, and obeys the triangle inequality.  We
use this function to compare the atom positions $u_i^a$ and $u_j^b$
even though they are defined from different supercells; only pathways
where all atoms move less than $\deltamax$ are considered.

Given our distance function, we calculate the distance $d(u_j^b + h
-u_i^a)$ for all $i$ and $j$ and shifts $h$.  We construct the
possible shifts $h$ for each $i$ and $j$ on a component-by-component
basis to minimize the number of $h$'s to consider.  If the
$l^\text{th}$ component of $u_j^b-u_i^a$ is less than 0, we use
$h_l=0$ or $h_l=1$; if instead it is greater than 0, we allow $h_l=0$
or $h_l=-1$.  In this way, we construct $2^3$ shifts $h$ for each pair
$i,j$.  This produces a table of $N$ by $8 N$ entries of distances.

For each atom $i$ we construct the set of atoms $C(i)=\{u_j^b + h\}$
where each atom is within $\deltamax$ of $u_i^a$.  These are the
allowed atom identifications for each $i$.  Our combinatorial problem
is on the sets $C(i)$; no two atoms $i$ and $i'$ may map to the same
atom $j$.  The set of possible pathways is found by iteration,
performed using recursion.  For $i=1$, we pick in turn one element of
$C(1)$; this will be $j(1)$.  For each $i>1$, we remove any entries of
$j(1)$ from $C(i)$; call the new sets $C_{(1)}(i)$.  We then pick in
turn one element of $C_{(1)}(2)$; this is $j(2)$.  For each $i>2$, we
remove any entries of $j(2)$ from $C_{(1)}(i)$; call the new sets
$C_{(1,2)}(i)$.  We repeat for successive $i$ until $i=N$, at which
point we have an entire mapping $j(i)$.  If at any point a set
$C_{(1,\ldots,k)}(i)$ for $i>k$ becomes empty or we exhaust all the
possible entries in a set we go back and make a different choice for
some smaller $i$.  This will produce all possible mappings consistent
with the initial sets $C(i)$.

After each possible pathway is constructed, we calculate the
$\delta_i$ and only keep pathways where $\sum_i \delta_i$ is zero.
Because we consider all possible $\dcom$ in our ``outer loop,'' we are
guaranteed to enumerate all possible pathways within our $\deltamax$
limit.  We store the possible pathways for each center of mass shift,
and then check that each pathway is unique.

\subsection{Uniqueness of pathways}
\label{ch:shuffles:unique}

Because each atom is indistinguishable, we define equivalence of
pathways in terms of the local environment that atoms will see along
the pathway.  We believe our definition for equivalence is sufficient,
but have not proven such.  This was tested by comparing against the
pathways produced in the common subgroup method.\cite{COMSUBS}

We use three tests to determine if two shuffle sets $\delta_i$ and
$\delta'_i$ are equivalent.  First, the sets of shuffle magnitudes
must be equal:
\begin{align*}
&\{(|[A]\delta_1|,|[B]\delta_1|),\ldots,(|[A]\delta_N|,|[B]\delta_N|)\}
=\\
&\{(|[A]\delta'_1|,|[B]\delta'_1|),\ldots,
  (|[A]\delta'_N|,|[B]\delta'_N|)\}.
\end{align*}
If this is true, then we next require that distance of closest
approach during the transformation---$\dca$---match.  We calculate
$\dca$ by simultaneously linearly interpolating all atoms from their
initial to final positions using a single variable $x$ while straining
the cell by $\mathbf{1}+x(\eps-\mathbf{1})$.  If the distance of
closest approach is the same, then we check the nearest neighbor
distance for each atom in the half-strained, half-shuffled
``intermediate'' supercell.  If those sets are the same, then we say
that we have two equivalent pathways.

These tests only comprise a set of {\em necessary} conditions, not
sufficient, for equivalence; this should not be seen as a severe
limitation.  First, two inequivalent paths marked as ``equivalent''
would be similar insofar as the transformation would produce similar
local environments for the atoms.  It's not entirely clear there would
be a very large difference in energy as the crystal transformed using
either pathway.  Second, our choice of tests has been checked by
comparing to an alternate method of atom identification using the
Wyckoff positions in the strained ``intermediate'' lattice; both
methods produced identical lists of unique pathways for many different
supercell choices.  While we don't know that the Wyckoff tests are not
also limited, they should not have the {\em same} limitations, and so
it is very likely that the limitations of each must be very small.
Finally, the real test of equivalence is that two pathways are
equivalent if for each atom in the first pathway, there is one and
only one atom in the second pathway that has the same local
environment during the transformation.  Such a test would require a
complicated $N!$ search, and would ultimately make the test for
uniqueness computationally intractable.  We believe our set of
conditions is a useful subset of the full equivalence test.

\subsection{Reduction by total shuffle magnitude and distance of
  closest approach}
\label{ch:method:rms}

We remove pathways with small distances of closest approach and large
root mean square (RMS) shuffle magnitudes because their energy
barriers will be high.  The majority of pathways generated for a given
$\deltamax$ are energetically unlikely because they require two atoms
to come very close together, or the majority of the atoms to move a
large distance.  For each supercell pair, we retain a subset of
pathways by examining their RMS shuffle magnitudes
\beu
\drms = \sqrt{\frac{1}{N}\sum_i^N d^2(\delta_i)}
 = \sqrt{\frac{1}{2N}\sum_i^N\left(
\left|[A]\delta_i\right|^2 + \left|[B]\delta_i\right|^2\right)},
\eeu
and the distance of closest approach for each.  We find the
\textit{largest} distance of closest approach for the supercell pair
$P([n], [m]; *)$, and set the minimum allowed distance of closest
approach to be $\min(\dca) = \max(\dca) - 0.1\text{~\AA}$.  For the
supercell pair, the maximum allowed RMS shuffle magnitude is
$\max(\drms) = \max\left( \sqrt{2}\min(\drms), \drms(\text{best } \dca
\text{ pathway})\right)$.

These two rules (a)~reject many of the poor candidates for each
supercell pair, (b)~while ensuring that at least one pathway for each
supercell pair is examined, even if it has a small distance of closest
approach.  If a supercell pair has a ``good'' solution (small $\drms$
and large $\dca$) we check only a few other possible pathways; if the
supercell pair does not have any ``good'' solutions, we check many
possible pathways for that pair.

\section{Calculating energy barriers}
\label{ch:method:barrier}

Once a set of pathways $P([\bar n], [\bar m][l]; u_i, \delta_i)$ is
known, we determine which pathways are the most probable for a given
material.  The sorting criterion we use is that the pathway with the
lowest energy barrier should be the most likely to occur during a
homogeneous transformation.  Most martensites are believed to
transform by heterogeneous nucleation, and there are some caveats
involved in translating the homogeneous pathway into a heterogeneous
pathway.  However, those limitations are material specific, and so
cannot be solved for the general case; these issues have been
addressed for titanium.\cite{Trinkle03}

Calculation of the energy barrier of a pathway requires (a) an atomic
interaction potential and (b) a method for determining the energy
barrier of a transformation.  The latter can be solved accurately with
some computational effort, or approximately with less effort.  Because
ultimately only the pathways with the lowest energy barriers are
important and not the many higher energy pathways, we use approximate
barrier calculations for the vast majority of the pathways.  This
leaves the accurate calculation of the barrier to a subset of the most
likely pathways, reducing computational effort.

We use three methods to calculate the energy barrier, from most
accurate and computationally intensive to least.  The first
method---the \term{nudged elastic band} method\cite{Jonsson98}---gives
the accurate transformation barrier.  The \term{landscape barrier}
uses a reduced phase space of one strain and one concerted shuffle
variable. The lack of relaxation makes the landscape barrier an
overestimate of the nudged elastic band barrier.  The \term{elastic
barrier} is an approximation to the landscape barrier, constructed
only from the strain.  The lack of atomic ``stiffness'' from phonons
makes the elastic barrier an underestimate of the landscape barrier.
Details of the landscape and elastic barriers have been published
elsewhere.\cite{TrinkleThesis}

\section{Titanium \atow\ results}

The pathway generation and sorting algorithm was applied to the Ti
\atow\ transformation, and a new pathway emerged with an energy
barrier lower than all others by over a factor of
four.\cite{Trinkle03}  Herein we provide a detailed study of the
methodology used and the remaining low energy barrier pathways
studied.

\subsection{Computational details}

Parameters for the pathway generation algorithm are chosen to ensure
sampling of all relevant pathways, and the total energies are computed
to yield accurate energy barriers.  The 2-atom unit cell of hcp and
3-atom unit cell of omega require that all supercell pairs contain a
multiple of 6 atoms; so we use $N=6$ and 12.  We require our diagonal
strains be less than $\epsmax=1.333$; for comparison, the Burgers
pathway for hcp to bcc requires a diagonal strain component of only
1.1.\cite{Burgers34} We keep only pathways with elastic barriers less
than $\Ecut=100\text{~meV/atom}$.  From those supercell pairs, we
construct pathways with a maximum shuffle magnitude $\deltamax$ less
than 2.0~\AA; this is comparable to the interatomic distance in
$\alpha$ and $\omega$, and, as seen later, is much larger than the
shuffles in our lowest energy pathway.  Each supercell pair produces
multiple pathways with shuffles smaller than this value.  As discussed
in Section~\ref{ch:method:rms}, we keep pathways with closest approach
distance $\dca$ within 0.1~\AA\ of the largest $\dca$.  We reject any
pathways with an RMS shuffle $\drms$ which is larger than both the
$\drms$ of the best $\dca$ pathway found and $\sqrt{2}$ of the
smallest $\drms$.  This gives us a set of pathways with good distance
of closest approach, RMS shuffle, and guarantees at least one pathway
for each supercell pair.

We calculate total energies for the landscape energy barriers using a
tight-binding (TB) model.  The TB calculations are performed using the
molecular dynamics code \ohmms\cite{ohmms} and use Mehl and
Papaconstantopoulos's functional form\cite{NRL96,NRL02} with
parameters refit to reproduce full-potential density-functional total
energies for hcp, bcc, fcc, omega, and sc to within 0.5~meV/atom (see
Appendix~B of [\onlinecite{TrinkleThesis}]).  For each mechanism, we
use a k-point set equivalent to a $12\x 12\x 9$ or $12\x 12\x 8$ grid
in the original hcp lattice.  This k-point mesh is converged to
within 0.1~meV/atom against a $18\x 18\x 12$ k-point mesh for the
three lowest energy barrier pathways.  A Fermi broadening of 63~meV
(5~mRyd) was used with this k-point mesh to ensure a smooth electronic
density of states.

We calculate total energies and forces for the nudged elastic band
barrier using carefully converged \abinit\ calculations, performed
with \vasp.\cite{Kresse93,Kresse96b}  \Vasp\ is a plane-wave based
code using ultra-soft Vanderbilt type
pseudopotentials\cite{Vanderbilt90} as supplied by Kresse and
Hafner.\cite{Kresse94}  The calculations were performed using the
generalized gradient approximation of Perdew and Wang.\cite{Perdew91}
We include $3p$ electrons in the valence band and use a plane-wave
kinetic-energy cutoff of 400~eV and a $7\x 7\x 7$ k-point mesh
to ensure energy convergence to within 1~meV/atom.  We relax the
atomic positions and the unit cell shape and volume until the atomic
forces are less than 20~meV/\AA\ and the stresses are smaller than
20~MPa.

\subsection{Generated pathways}
\label{ch:results:pathways}

Table~\ref{tab:h2o:sorting} summarizes the resulting number of
supercell pairs and pathways generated and the cutoffs made at each
step in the algorithm.  The nudged elastic band energy barrier for the
final seven pathways is calculated, and the pathway with the lowest
barrier is TAO-1.\cite{Trinkle03}

\newcommand*{\markpathway}[1]{\textbf{#1}}
\newcommand*{\mypara}[1]{#1}

\begin{table}
\begin{center}
\begin{tabular}{lcccc}
&\multicolumn{2}{c}{\hrulefill\ 6 atom \hrulefill}
&\multicolumn{2}{c}{\hrulefill\ 12 atom \hrulefill}\\
\multicolumn{1}{c}{Sorting}
 &supercell& &supercell \\
\multicolumn{1}{c}{\underbar{step:}}
&\underbar{pairs}&\underbar{pathways}
&\underbar{pairs}&\underbar{pathways}\\
\hline
\mypara{1. $\epsmax$}&
  6 & -- & 128 & -- \\
\mypara{2. $\Ecut$}&
  4 & -- & 56 & -- \\
\mypara{3. $\deltamax$}&
  4 & \markpathway{55} & 56 & \markpathway{3790} \\
\mypara{4. $\dca$, $\drms$}&
  4 & \markpathway{6} & 56 & \markpathway{971} \\
\mypara{5. $E_\text{landscape}$}&
  2 & \markpathway{2} & 3 & \markpathway{5}
\end{tabular}
\end{center}

\caption{Number of supercell pairs, pathways generated, and cutoff
parameters in Ti hcp to omega pathway search.  Each consecutive step
builds on the selection made in the previous step: \textbf{1.}~We
generate an initial set of 134 supercell pairs with a given $\epsmax$
cutoff of 1.333.  \textbf{2.}~We choose the 60 supercell pairs that
have an elastic barrier less than $\Ecut$ of 100~meV/atom.
\textbf{3.}~The possible shuffles for each supercell pair within
$\deltamax$ of 2~\AA\ are determined to give 3,845 possible pathways.
\textbf{4.}~For each supercell pair, we define a minimum allowed
distance of closest approach ($\dca$), and a maximum root mean square
(RMS) shuffle magnitude ($\drms$), and keep the 977 pathways within
those limits.  \textbf{5.}~Finally, the 7 pathways with tight-binding
(TB) landscape barriers less than 90~meV/atom are kept, and their true
energy barriers calculated.}
\label{tab:h2o:sorting}
\end{table}

We generate an initial set of 134 supercell pairs with diagonal strain
components less than $\epsmax$ and reduce it to 60 pairs with
elastic barriers less than $\Ecut$ of 100~meV/atom.  The
three-dimensional surface of possible diagonal strain values
$\bar\eps_i$ satisfying $\inv{\epsmax}\le \bar\eps_i \le \epsmax$ and
$\bar\eps_1\bar\eps_2\bar\eps_3 = 1$ has sharp corners at the boundary;
because of this, we generate more supercell pairs than are used to
ensure that we produce all supercell pairs less than our elastic
barrier cutoff. This procedure is analogous to inscribing a circle
inside of a regular hexagon.

Using larger supercells fails to produce any supercell pairs with
strains smaller than the smallest strains already found.  The
possibility for better pathways for larger supercell sizes is unlikely
as there are no 18- or 24-atom supercell pairs found with strains
smaller than the smallest strain found using 12-atom supercells.  We
keep the 12-atom supercell pairs because there \textit{is} a 12-atom
supercell pair that has smaller strains than the smallest strain for a
6-atom pathway; three of the seven low energy pathways have this
smallest strain value.~\footnote{It is worth noting that the lowest
energy pathway does have a larger strain than the smallest strain
found, but it is also a 6-atom pathway.}

The 60 supercell pairs generate 55 6-atom pathways and 3790 12-atom
pathways with $\deltamax$ of 2.0~\AA; this set is reduced to 6 6-atom
and 971 12-atom pathways using the closest approach distances and RMS
shuffle magnitudes as described in Section~\ref{ch:method:rms}.  The
reduction for the 6-atom pathways is more severe because each of the
supercell pairs has a ``good'' solution (small RMS shuffle magnitude
and large closest approach distance) so we examine fewer alternative
pathways.  Many of the 12-atom supercell pairs do not have ``good''
solutions, so we examine more alternative pathways.  A check of this
reduction for good solutions was performed by calculating the
landscape barrier of the ``best'' excluded pathway for the TAO-1
6-atom supercell pair; it has a TB landscape barrier of 160~meV/atom,
compared to 43~meV/atom for TAO-1.

The 977 pathways are then reduced to seven: three with landscape
barriers less than 80 meV, and four with landscape barriers between 80
meV and 90 meV.  The three lowest pathways are TAO-1 (6-atom), TAO-2
(6-atom), and Silcock (12-atom); the latter is a pathway already
proposed in the literature.\cite{Silcock58}  The remaining six
pathways are new; though TAO-1 and TAO-2 are related to Usikov and
Zilberstein's $\alpha$$\to$$\beta$$\to$$\omega$ pathway through an
intermediate bcc structure.\cite{Usikov73}  We calculate the nudged
elastic band barrier for Silcock and TAO-1 through TAO-6.

\begingroup
\begin{table*}
\begin{tabular} {l@{\quad}l@{\quad}l|rccc}
\hline\hline &&&\multispan4{\hfill Wyckoff positions \hfill}\\
Path&Subgroup&\multispan1{\hfill Supercell lattice\hfill\vline}&
$m$ & label & $\alpha$ & $\omega$ \\
\hline\hline
TAO-1 &15 $C2/c$&$\alpha: (-1,-1,2),(1,-1,0),(1,1,1)$&
               2&$(e)$&$0,7/12,1/4$&$0,1/2,1/4$\\
(6-atom)       &&$\omega: (-2,-1,2),(0,-1,0),(0,0,2)$&
               4&$(f)$&$1/3,5/12,1/12$&$1/3,1/2,1/6$\\
\hline
TAO-2 &11 $P2_1/m$&$\alpha: (1,0,0),(0,0,1),(0,-3,0)$&
               2&$(e)$&$1/3,1/4,7/9$&$ 1/2,1/4,3/4$\\
(6-atom)       &&$\omega: (0,0,-1),(-1,0,0),(1,2,-2)$&
               2&$(e)$&$1/3,1/4,1/9$&$1/3,1/4,1/12$\\
               &&&
               2&$(e)$&$1/3,1/4,4/9$&$2/3,1/4,5/12$\\
\hline
Silcock &57 $Pbcm$&$\alpha: (-1,0,0),(-3,-6,0),(0,0,1)$&
               4&$(d)$&$3/4,23/36,1/4$&$1/2,5/8,1/4$\\
(12-atom)      &&$\omega: (0,0,1),(2,4,0),(-1,0,0)$&
               4&$(d)$&$3/4,11/36,1/4$&$1,7/24,1/4$\\
               &&&
               4&$(d)$&$3/4,35/36,1/4$&$1,23/24,1/4$\\
\hline
TAO-3 &6 $Pm$   &$\alpha: (1,0,0),(0,0,1),(0,-6,0)$&
               1&$(a)$&$1/3,0,1/18$&$0,0,0$\\
(12-atom)      &&$\omega: (0,0,-1),(-1,0,0),(2,4,-3)$&
               1&$(a)$&$1/3,0,2/9$&$0,0,1/6$\\
               &&&
               1&$(a)$&$1/3,0,7/18$&$1/2,0,1/3$\\
               &&&
               1&$(a)$&$1/3,0,5/9$&$1/2,0,1/2$\\
               &&&
               1&$(a)$&$1/3,0,13/18$&$1/2,0,2/3$\\
               &&&
               1&$(a)$&$1/3,0,8/9$&$0,0,5/6$\\
               &&&
               1&$(b)$&$2/3,1/2,1/9$&$1/4,1/2,1/12$\\
               &&&
               1&$(b)$&$2/3,1/2,5/18$&$1/4,1/2,1/4$\\
               &&&
               1&$(b)$&$2/3,1/2,4/9$&$5/4,1/2,5/12$\\
               &&&
               1&$(b)$&$2/3,1/2,11/18$&$3/4,1/2,7/12$\\
               &&&
               1&$(b)$&$2/3,1/2,7/9$&$3/4,1/2,3/4$\\
               &&&
               1&$(b)$&$2/3,1/2,17/18$&$3/4,1/2,11/12$\\
\hline
TAO-4 &11 $P2_1/m$&$\alpha: (1,0,0),(0,0,1),(0,-6,0)$&
               2&$(e)$&$1/3,1/4,13/18$&$-1/8,1/4,17/24$\\
(12-atom)      &&$\omega: (0,0,-1),(-1,0,0),(2,4,-3)$&
               2&$(e)$&$1/3,1/4,2/9$&$3/8,1/4,5/24$\\
               &&&
               2&$(e)$&$1/3,1/4,8/9$&$-1/8,1/4,7/8$\\
               &&&
               2&$(e)$&$1/3,1/4,1/18$&$7/8,1/4,1/244$\\
               &&&
               2&$(e)$&$1/3,1/4,7/18$&$3/8,1/4,3/8$\\
               &&&
               2&$(e)$&$1/3,1/4,5/9$&$3/8,1/4,13/24$\\
\hline
TAO-5 &6 $Pm$&$\alpha: (-1,0,0),(0,0,2),(0,3,0)$&
           1&$(a)$&$2/3,0,2/9$&$5/6,0,1/4$\\
(12-atom)  &&$\omega: (0,0,1),(-2,0,0),(-1,-2,-1)$&
           1&$(a)$&$2/3,0,5/9$&$0,0,7/12$\\
           &&&
           1&$(a)$&$2/3,0,8/9$&$2/3,0,11/12$\\
           &&&
           1&$(b)$&$2/3,1/2,2/9$&$5/6,1/2,1/4$\\
           &&&
           1&$(b)$&$2/3,1/2,5/9$&$1,1/2,7/12$\\
           &&&
           1&$(b)$&$2/3,1/2,8/9$&$2/3,1/2,11/12$\\
           &&&
           2&$(c)$&$1/3,1/4,1/9$&$1/2,1/4,1/12$\\
           &&&
           2&$(c)$&$1/3,1/4,4/9$&$1/6,1/4,5/12$\\
           &&&
           2&$(c)$&$1/3,1/4,7/9$&$1/3,1/4,3/4$\\
\hline
TAO-6 &11 $P2_1/m$&$\alpha: (1,0,0),(0,0,1),(0,-6,0)$&
               2&$(e)$&$1/3,1/4,13/18$&$23/24,1/4,17/24$\\
(12-atom)      &&$\omega: (0,0,-1),(-1,0,0),(2,4,-5)$&
               2&$(e)$&$1/3,1/4,2/9$&$11/24,1/4,5/24$\\
               &&&
               2&$(e)$&$1/3,1/4,8/9$&$5/8,1/4,7/8$\\
               &&&
               2&$(e)$&$1/3,1/4,1/18$&$7/24,1/4,1/24$\\
               &&&
               2&$(e)$&$1/3,1/4,7/18$&$1/8,1/4,3/84$\\
               &&&
               2&$(e)$&$1/3,1/4,5/9$&$19/24,1/4,13/24$\\
\hline\hline
\end{tabular}
\caption{Pathways for the \atow\ phase transition in
Ti. Each pathway is defined by a maximal common subgroup of both the
$\alpha$ and $\omega$ phases and Wyckoff positions.  A supercell
containing both phases is given in terms of the lattice vectors of
$\alpha$ and $\omega$ unit cells; the lattice vectors determine $[n]$
and $[m]$.  The atomic positions for each phase are given in the
setting of the common subgroup; together the positions define $u_j$
and $\delta_j$.  The multiplicity $m$ for each Wyckoff position is given
with the label of the position.}
\label{tab:symmetry}
\end{table*}

Table~\ref{tab:symmetry} gives details for the seven lowest-barrier
pathways of interest.  Along each pathway, the symmetry is a subgroup
of both crystals \crys{a} and \crys{b}. Using \comsubs,\cite{COMSUBS}
we obtained the maximal symmetry of this subgroup for each pathway.
In the table, we list this maximal common subgroup along with atomic
positions in the setting of this subgroup.  This gives a complete
description of the mapping of atoms in \crys{a} onto atoms in
\crys{b}.  For example, in the TAO-1 pathway, the symmetry of the
crystal is monoclinic (space group 15 $C2/C$) as it transforms from
$\alpha$ to $\omega$.  The atoms are at the Wyckoff $(e)$ and $(f)$
sites of that space group.  During the transformation the atoms at the
$(e)$ site $(0,y,1/4)$ move from $(0,7/12,1/4)$ to $(0,1/2,1/4)$ and
the atoms at the $(f)$ site $(x,y,z)$ move from $(1/3,5/12,1/2)$ to
$(1/3,1/2,1/6)$.

It is important to note that in the nudged-elastic band method, the
atomic positions are relaxed under {\em triclinic symmetry}: no point
symmetries are assumed or enforced.  We find the resulting symmetry
along each pathway to be nonetheless equal to the maximal symmetry
given in Table~\ref{tab:symmetry}.  In each of these cases, we
conclude that the lowest barriers occur for pathways with maximal
symmetry.

Table~\ref{tab:results:NEB} summarizes the energy barriers for the
seven lowest pathways of interest.  The TB NEB barrier is bounded
above by the landscape barrier and below by the elastic barrier.  The
difference between the TB NEB barrier and the \abinit\ NEB barrier is
due primarily to the latter's cell size and shape relaxations.

\begin{table}
\begin{center}
\begin{tabular}{lccccccc}
 & \entry{Silcock} & \entry{TAO-1} & \entry{~2~} & \entry{~3~} &
 \entry{~4~} & \entry{~5~} & \entry{~6~}\\
\multicolumn{8}{c}{Homogeneous barriers (in meV/atom)}\\
Elastic:     & 3.7 & 18  & 21  & 3.7 & 3.7 & 21 & 74 \\
TB Landscape:& 60  & 43  & 61  & 83  & 80  & 84 & 81 \\
TB NEB:      & 54  & 24  & 52  & --  & --  & -- & -- \\
\Abinit\ NEB:& 31  &  9  & 58  & 32  & 68  & 37 & 69 \\
\multicolumn{8}{c}{Transformation information}\\
Supercell size:  & 12 & 6 & 6  & 12  & 12  & 12 & 12 \\
Orientation & II & I & II & II & II & II & II \\
\quad Relations: 
&\multicolumn{7}{l}{I: $(0001)_\alpha \;\|\; (0\bar 111)_\omega$, 
$[11\bar 20]_\alpha  \;\|\; [01\bar 11]_\omega$} \\
&\multicolumn{7}{l}{II: $(0001)_\alpha \;\|\; (11\bar 20)_\omega$, 
$[11\bar 20]_\alpha  \;\|\; [0001]_\omega$}
\end{tabular}
\end{center}

\caption{Comparison of lowest landscape barrier pathways. Energy
barriers: Four different methods for calculating the energy barrier
for the three pathways are shown, from least accurate to most
accurate.  The elastic barrier only accounts for the strain in each
pathway.  The landscape barrier uses a simple combined shuffle for
each, and a tight-binding total energy.  Finally, the NEB calculation
is done with the tight-binding method for the lowest three and
\textit{ab~initio} for all seven to accurately determine the barrier.
Orientation Relations: The relative orientation between $\alpha$ and
$\omega$ is shown for each pathway.  N.B.: Silcock and the TAO-2
through TAO-6 pathways have the same orientation relations.}
\label{tab:results:NEB}
\end{table}

Figure~\ref{fig:results:EnBarrier} shows the energy barrier for the
seven lowest energy barrier pathways.  The plots are grouped according
to the supercell pairs in each pathway.  The difference in energy
barrier among pathways with the same supercell shows the importance of
theory in understanding the microscopic mechanism---measurements of
the final $\omega$ orientation relative to the $\alpha$ can only rule
out possible pathways.

\begin{figure}
\begin{center}
\includegraphics[width=3.3in]{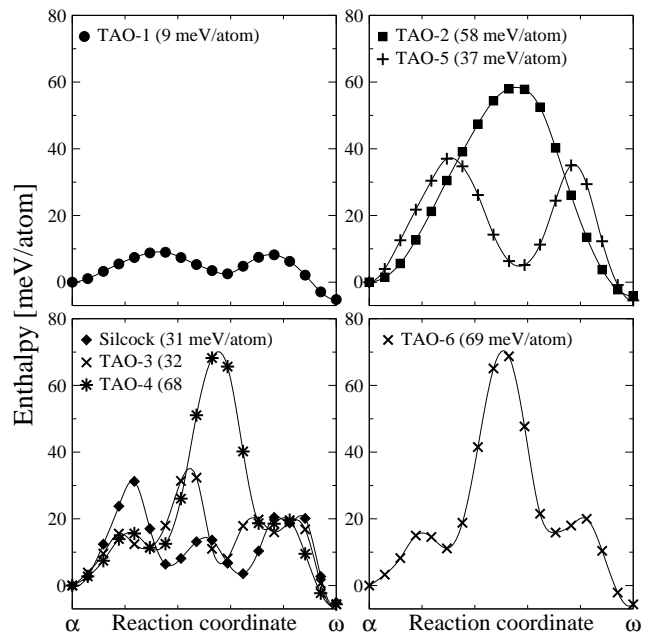}
\end{center}

\caption{Energy barrier at zero pressure for the seven lowest 
energy pathways, using \NEBVASP\ at 16 intermediate states.  The
pathways are grouped by their strains; each set have related
supercell pairs.  The effect of different shuffles on the energy
barrier can be seen by comparing TAO-2 to TAO-5, and Silcock, TAO-3,
and TAO-4 to each other.  The latter three pathways have the smallest
amount of strain, but the homogeneous barrier is smallest for TAO-1.}
\label{fig:results:EnBarrier}
\end{figure}

Figure~\ref{fig:results:TAO1-simple} illustrates a simple geometric
picture of the low barrier TAO-1 pathway.  The basal plane of $\alpha$
is a series of hexagons surrounding individual atoms; these hexagons
then break into two 3-atom pieces.  Each 3-atom piece swings in
opposite directions out of the basal plane, and connects with a mated
3-atom piece above or below.  These form the honeycomb sublattice of
$\omega$; the remaining ``unmoved'' atoms form the $A$ sublattice in
$\omega$.  This transformation transforms the $(0001)$ plane of
$\alpha$ into the $(01\bar11)$ plane of $\omega$.

\begin{figure}

\centerline{
\epsfig{file=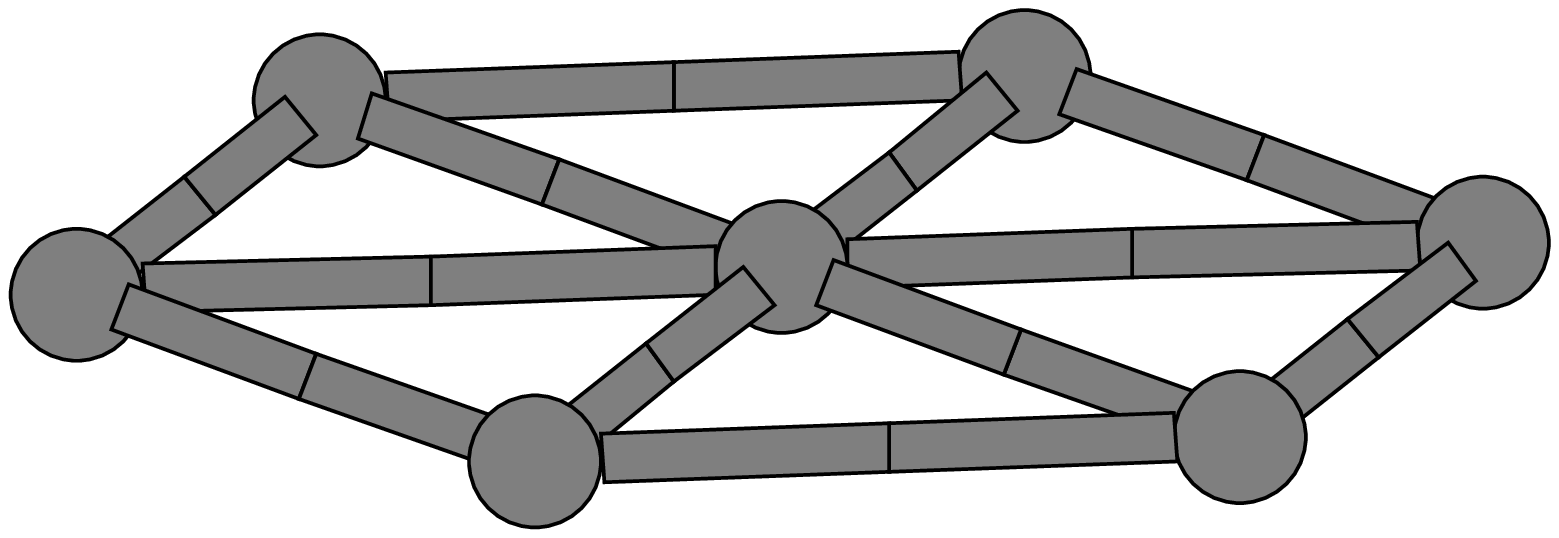, width=1.4in}
\raise0.5in\hbox{\quad$\Longrightarrow$\quad}
\epsfig{file=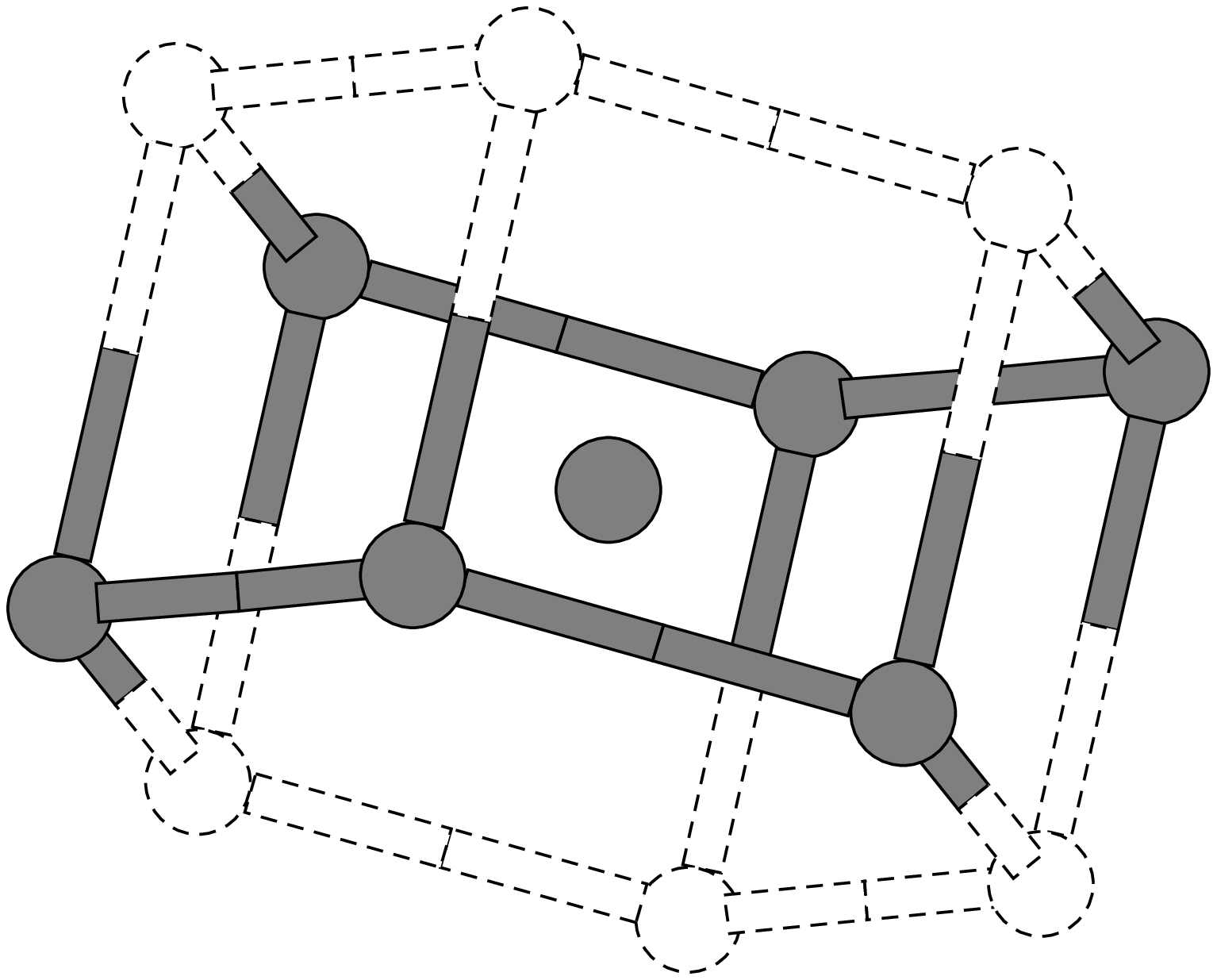, width=1.4in}
}

\centerline{
\hbox to 2in{\hfil$(0001)$-$\alpha$ plane\quad}
\hbox{\quad$\Longrightarrow$\quad}
\hbox to 2in{\quad$(01\bar11)$-$\omega$ plane\hfil}
}

\caption{TAO-1 acting on a single $\alpha$ basal plane.  A hexagon in
the basal plane is broken into two 3-atom pieces.  Each 3-atom piece
swings out of the basal plane in opposite directions, to create half
of the honeycomb lattice in two parallel $(0001)$ planes of $\omega$.
The remaining atom in the center forms the $A$ sublattice lying in the
$(0001)$ planes of $\omega$.  The next $\alpha$ basal layer creates
the dashed atoms in the $\omega$ honeycomb.  This transformation turns
the $(0001)$ plane of $\alpha$ into the $(01\bar11)$ plane of
$\omega$.}
\label{fig:results:TAO1-simple}
\end{figure}

Of the remaining six pathways, TAO-2 and Silcock are both related to
existing pathways in the literature, while TAO-3 through TAO-6 are
related to the other three pathways.  TAO-2 has the same endpoints as
Usikov and Zilberstein's variant II pathway,\cite{Usikov73} but avoids
the intermediate bcc phase.  The Silcock pathway is identical to the
original published pathway.\cite{Silcock58} Both TAO-3 and TAO-4 use
the Silcock supercell pairing, though with different shuffles.  TAO-5
is a cell doubling of the TAO-2 supercell pairing with different
shuffles.  Finally, TAO-6 uses the same $\alpha$ supercell as Silcock,
but connects it to a different $\omega$ supercell.

TAO-1 has the lowest energy barrier that best trades off between
shuffle and strain.  During the transformation, the closest nearest
neighbor distance is 2.63~\AA, which is larger than the 2.55~\AA\
value for TAO-2, 2.58~\AA\ for Silcock, TAO-3, TAO-4, and TAO-6.  Only
TAO-5 has a comparable closest nearest neighbor distance of 2.63~\AA;
this larger distance helps to explain the lower barrier compared to
TAO-2 with the same strains.  Because the nearest-neighbor distances
in $\alpha$ and $\omega$ are 2.93~\AA\ and 2.65~\AA, respectively, it
is not surprising that TAO-1 has the lowest barrier.  The larger
barrier of TAO-5 compared to TAO-1 may be due to more complicated
geometric changes during the pathway.  The lack of a simple, single
criterion to explain the difference in energy barriers shows the
necessity of \abinit\ studies.

\section{Conclusion}

We have presented a general systematic \pgs\ algorithm for martensitic
transformations.  The method enumerates all possible pathways within a
few geometric restrictions.  When applied to the Ti \atow\
transformation, a previously unknown pathway emerges with a barrier
much lower than all other pathways.  Geometric criteria are useful in
reducing the large set of possible pathways to a more manageable set,
but \abinit\ studies are required to ultimately find the lowest energy
barrier pathway.

\begin{acknowledgments}
We thank J.~W.~Wilkins for helpful discussions.  DRT thanks Los Alamos
National Laboratory for its hospitality and was supported by a Fowler
Fellowship at Ohio State University. This research is supported by DOE
grants DE-FG02-99ER45795 (OSU), DE-FG02-03ER46059 (BYU), and
W-7405-ENG-36 (LANL). Computational resources were provided by the
Ohio Supercomputing Center and NERSC.
\end{acknowledgments}

\end{document}